\documentclass[nohyper,notoc]{JHEP}

\usepackage{amsmath,euscript,array,amssymb,cite,bm} 
\def\cN{{\cal N}}

\def\dalpha{{\dot\alpha}}

\def\sigmabar{\bar\sigma}

\def\Tr{{\rm Tr}}

\def\SU{\text{SU}}

\def\Dbarslash{\,\,{\raise.15ex\hbox{/}\mkern-12mu {\bar\D}}}
\def\Dslash{\,\,{\raise.15ex\hbox{/}\mkern-12mu \D}}
\def\delslash{\,\,{\raise.15ex\hbox{/}\mkern-9mu \partial}}
\def\delbarslash{\,\,{\raise.15ex\hbox{/}\mkern-9mu {\bar\partial}}}

\def\hf{{\textstyle{1\over2}}}

\def\l{\lambda} 
\def\s{\sigma}

\def\D{{\cal D}}
\def\Dbarslash{\,\,{\raise.15ex\hbox{/}\mkern-12mu {\bar\D}}}
\def\delslash{\,\,{\raise.15ex\hbox{/}\mkern-9mu \partial}}
\def\Dslash{\,\,{\raise.15ex\hbox{/}\mkern-12mu \D}}
\def\Skinst{S^k_{\rm inst}}

\def\adss{AdS_5\times S^5}
\def\ads{AdS_5}
\def\sqrtwo{\sqrt{2}\,}

\newcommand{\SO}{{SO}}

\def\bigC{{\rm l}\!\!\!{\rm C}}
\def\hg{\hat{\gamma}}
\def\={\, =\, }
\def\+{\, +\, }
\def\-{\, -\, }
\def\cQ{{\cal Q}}

\newcommand{\be}{\begin{equation}}
\newcommand{\ee}{\end{equation}}
\def\bea{\begin{eqnarray}}
\def\eea{\end{eqnarray}}

\newcommand{\ft}[2]{{\textstyle\frac{#1}{#2}}}

\def\bSYM{$\beta$-SYM}

\title{Instanton test of non-supersymmetric deformations of the $\bm{ AdS_5\times S^5}$}

\author{Callum Durnford, George Georgiou and Valentin V.~Khoze\\
Department of Physics and IPPP, University of Durham,
Durham, DH1 3LE, UK\\
E-mails: {\tt callum.durnford, george.georgiou, valya.khoze@durham.ac.uk}}

\abstract{We consider instanton effects in a non-supersymmetric gauge theory obtained
by marginal deformations of the $\cN=4$ SYM. This gauge theory is 
expected to be dual to type IIB string theory on the $AdS_5$ times deformed-$S^5$ background.
From an instanton calculation in the deformed gauge theory we extract the prediction
for the dilaton-axion field $\tau$ in dual string theory. In the limit of small deformations
where the supergravity regime is valid, our instanton result 
reproduces the expression for $\tau$ of the supergravity solution found by Frolov. 
 }

\preprint{{\tt hep-th/0606111}\\IPPP/06/35\\
DCPT/06/70}

\begin{document}

\section{Introduction and discussion of results} 

Much has been learnt recently about the gauge theory -- string theory duality by investigating
how the AdS/CFT correspondence \cite{Maldacena}
is realised when the $\cN=4$ supersymmetric gauge theory is deformed by exactly marginal
operators 
\cite{LM,LS,BL,Ah2,DHK,Frolov,FRT1,BR,FRT,FG,PSZ,Rossi,MPSZ,GK,CK,MPPSZ}. 
Since the gauge theory stays conformal it is expected
to be dual (in the appropriate limit) to a supergravity solution with Anti-de Sitter
geometry. There are two points which make these marginal deformations particularly interesting.
First, is that these deformations give a continuous family of theories parameterised by the deformation
parameters $\beta_i$. The AdS/CFT duality provides a mapping between a gauge theory and a string theory 
for each value of $\beta_i$.
By studying the $\beta$-dependence in gauge theory and in the dual supergravity
(or string theory) one thus gets a more detailed understanding of the AdS/CFT correspondence.
The second feature of marginal $\beta$-deformations is that they break (partially or completely) 
the supersymmetry of the 
original $\cN=4$ theory. 

Lunin and Maldacena \cite{LM} have found a supergravity dual of the 
$\beta$-deformed $\cN=4$ super Yang-Mills
theory (\bSYM) which preserves $\cN=1$ supersymmetry. In a recent paper \cite{GK} two of the present authors
have tested this supergravity solution and the resulting string theory effective action against 
an instanton calculation on the gauge theory side. It was found in \cite{GK} that the correct expression
for the dilaton-axion supergravity field $\tau$ 
was reproduced by instanton effects in gauge theory, and that the higher-derivative terms in the 
string theory effective action included the appropriate modular forms $f_n(\tau,\bar\tau)$ of this
$\tau$ as one would expect from the $SL(2,Z)$ duality in IIB string theory.

One way of realising the solution generating method in
\cite{LM} is via a combined T-duality-shift-T-duality (TsT) transformation of
the supergravity $\adss$ geometry. This approach enabled Frolov \cite{Frolov}
to extend the method and 
to find a three-parameter family of non-supersymmetric supergravity
solutions. This background has to be AdS/CFT dual to a 
non-supersymmetric conformal gauge theory obtained by a certain three-parameter deformation of
the $\cN=4$ SYM.

In this paper we apply the instanton approach of Refs. \cite{MO3,GK} to
investigate this non-supersymmetric gauge theory and to test
the supergravity solution of Ref. \cite{Frolov}. 
In section \textbf{2} we write down the supergravity solution 
of \cite{Frolov} parameterised by three real deformations $\gamma_i$ and
specify the corresponding $\gamma_i$-deformed gauge theory. 
In section \textbf{3} we carry out an instanton calculation in the
$\gamma_i$-deformed gauge theory with a view to reconstruct the dilaton-axion
supergravity field $\tau$ from gauge theory.
In the appropriate double-scaling limit, $\gamma_i \ll 1,$ our result 
\be
\label{OIsugrares}
\tau \, =\, \tau_0 + {2N\pi i}\left( \gamma_3^2 \, \mu_1^2\mu_2^2
  \, +\, \gamma_1^2 \, \mu_2^2\mu_3^2
  \, +\, \gamma_2^2 \, \mu_3^2\mu_1^2 \right)
\ee
reproduces
the $\tau$-field of Frolov's supergravity dual. Here $\tau_0$ is the usual complexified coupling constant
in gauge theory, $\gamma_i$ are the three deformation parameters, and $\mu_i$ are coordinates
on the deformed $S^5$ sphere in supergravity.  
In section \textbf{4} we generalise our set-up to include complex-valued
deformations $\beta_i=\gamma_i +i \sigma_i.$ Our main results there are Eqs.~\eqref{Fcom}-\eqref{fin-al}.

The fact that instanton contributions in gauge theory confirm the non-supersymmetric
supergravity solution of Ref.~\cite{Frolov} is our main result. Both expressions,
in gauge theory and in supergravity, are continuous functions of the three complex deformation parameters.
What is interesting about this matching
is not merely the fact that there is a non-trivial agreement between gauge theory and supergravity, 
but also that the Yang-Mills instanton calculation which is intrinsically valid only at weak coupling,
$g^2 N \ll 1,$ $N\to \infty$,
appears to give the correct result in the strong coupling limit,
$g^2 N \gg 1,$ relevant for comparison with the 
supergravity.
This agreement between the strong and the weak coupling limits
is completely analogous to the previously known instanton tests of
AdS/CFT correspondence in the $\cN=4$ SYM context in Refs.~\cite{BG,BGKR,DKMV,DHKMV,MO3,GreenKovacs,Review}
and more recently in the context of
supersymmetry-preserving $\beta$-deformations in Ref.~\cite{GK}. 
In all known cases, leading order contributions of Yang-Mills instantons calculated at
$g^2 N \ll 1,$ 
match with contributions of D-instantons in supergravity in the opposite limit $g^2 N \gg 1.$
The agreement holds only for the instanton part of the answer, it is known that
perturbative contributions in gauge theory and in string theory do not match.
This suggests that there should exist a non-renormalisation theorem which
would apply
to the instanton effects and explain the agreement. We refer the reader to 
Refs.~\cite{MO3,GoGreen} and \cite{GK,Review} for a more detailed discussion on this point.

In this paper we find that the agreement in the instanton sector
persists in the non-supersymmetric case. This implies that the non-renormalisation theorem
is not dictated by supersymmetry. We expect that the origin of the agreement lies in identifying 
Yang-Mills instantons with D-instantons as the `extended' objects or defects in both theories.

\section{Three-parameter deformation of the $\bm{\adss}$}

We begin by reviewing the theories on each side of the gauge/string duality we
wish to study. The solution generating tool on the supergravity side is the combination of
T-dualities and coordinate shifts known as a TsT transformation. These allow one
to start with the known duality between IIB supergravity on a flat background
and $\cN=4$ SYM, and generate new supergravity backgrounds \cite{LM,Frolov}. The deformation on
the gauge theory side will be incorporated by introducing an appropriate star-product between
fundamental fields. For the most part we will concern ourselves with real valued
deformations of the theory. The issues which arise for complex deformations will
be discussed in section \textbf{4}.

\subsection{Supergravity dual} 

In order to perform supergravity TsT transformations one must first identify
suitable tori in the initial geometry. In the case of \cite{LM} this torus was
chosen to be the one dual to the $U(1)\times U(1)$ global symmetry of 
\bSYM. If we parameterise this torus with angular variables
$(\varphi_1,\varphi_2)$, then a TsT transformation with parameter $\hg$ is the following: T-dualise
in the $\varphi_1$ direction, perform the shift $\varphi_2 \rightarrow \varphi_2
+ \hg \varphi_1$, T-dualise again along $\varphi_1$. The resulting supergravity
solution was shown in \cite{LM} to be dual to $\beta$-SYM for small, real $\beta$
under the association $\hg = R^2 \beta$ where $R$ is the radius of $S^5$.

The $S^5$ factor of $\adss$ can be parameterised with the
coordinates $\mu_1$, $\mu_2$, $\mu_3$ with $0\leq \mu_i\leq 1$ subject to
$\mu_1^2+\mu_2^2+\mu_3^2=1$ and the angular coordinates $\phi_1$, $\phi_2$,
$\phi_3$. There are clearly three independent choices of torus corresponding to
the pairs $(\phi_1,\phi_2), (\phi_2,\phi_3)$ and $(\phi_1,\phi_2)$. The three
parameter deformation constructed in Ref.~\cite{Frolov} follows by performing a separate TsT
transformation on each of these, with shift parameters $\hg_3$,
$\hg_1$ and $\hg_2$ respectively. The resulting type IIB supergravity background of Frolov
written in string frame with $\alpha'=1$ takes the form \cite{Frolov}:
\bea
\label{threepar}
ds^2_{{\rm{ str}}} &=& R^2\left[ ds^2_{{\rm{ AdS}}} + \sum_i\left(d\mu_i^2\+ G\, \mu_i^2\, d\phi_i^2 \right) \+
G\, \mu_1^2\,\mu_2^2\,\mu_3^2\left( \sum_i \hg_i\,d\phi_i\right)^2\right]\, ,\\ \nonumber
G^{-1}&=& 1 \+ \hg_3^2\,\mu_1^2\,\mu_2^2\+ \hg_1^2\,\mu_2^2\,\mu_3^2\+\hg_2^2\,\mu_3^2\,\mu_1^2\, ,\qquad
e^{2\phi}=e^{2\phi_0}\,G\, , \\ \nonumber
B^{NS}&=&R^2 G\left( \hg_3\, \mu_1^2\,\mu_2^2\,d\phi_1\wedge d\phi_2
\+ \hg_1\, \mu_2^2\,\mu_3^2\,d\phi_2\wedge d\phi_3\+\hg_2\,
\mu_3^2\,\mu_1^2\,d\phi_3\wedge d\phi_1\right)\,
\eea
We present here only the fields that will be relevant for our
purposes. The
full complement, including the RR forms $C_2$ and $C_4$ and self-dual five-form fields is given
  in \cite{Frolov}. To make contact with the dual gauge theory we have the usual
  AdS/CFT relation $R^4\equiv 4\pi e^{\phi_0}N=\sqrt{\l}$.
The real deformation parameters $\hg_i$ appearing in \eqref{threepar} are related to the $\gamma_i$ deformations
on the gauge theory side via a simple rescaling,  
 $\hg_i = R^2\gamma_i.$
 We note that the dilaton field $\phi$ in \eqref{threepar} 
 is not simply a constant, but depends on the coordinates of the deformed sphere $\tilde{S}^5.$
 (The axion field $C=C^0$ is a constant for real-valued deformations $\gamma_i$.)
 
When all three deformation parameters are equal,
$\hg_1\,=\,\hg_2\,=\,\hg_3\,\equiv\,\hg$, this solution reverts to that of
Lunin and Maldacena \cite{LM}, and the dual gauge theory is \bSYM.

\subsection{Gauge theory formulation}

The Frolov's supergravity solution \eqref{threepar} 
with three real deformations $\gamma_i$ contains the
$\ads$ factor.
Thus it is expected to be dual to a conformal gauge theory
obtained by exactly marginal but non-supersymmetric deformations of the $\cN=4$ SYM.
More precisely, the gauge theory should be conformal in the large number of colours limit
(which we always assume in this paper) where the supergravity approximation to string theory 
can be trusted.

We will be considering non-supersymmetric deformations of the $\cN=4$ gauge theory, parameterised
by three phases, $e^{i\pi\gamma_1},$ $e^{i\pi\gamma_2}$ and $e^{i\pi\gamma_3},$ with 
real parameters $\gamma_i.$ 
To ensure conformal invariance of the theory in the large $N$ limit, it is convenient to introduce these 
phase-deformations via a star-product approach.\footnote{Below we will use the fact \cite{Filk} that associative
star-products do not change planar diagrams of the original $\cN=4$ SYM. This means that in the large $N$ limit
the resulting deformed gauge theory will remain conformal at least in perturbation theory.}
We take the component Lagrangian of the
$\cN=4$ supersymmetric Yang-Mills and modify all products of fields there into star-products.
For any pair of fields $f$ and $g$, the star-product which gives rise to our deformations is \cite{FRT}:
\be
f * g \,\equiv\, e^{-i\pi\, Q^f_i Q^g_j \, \epsilon_{ijk}\gamma_k}\, f g
\label{stardef}
\ee
Here $Q^f_i$ and $Q^g_i$ are the charges of the fields $f$ and $g$ under the $i=1,2,3$ Cartan generators 
of the $SU(4)_R$ R-symmetry of the original $\cN=4$ SYM. The values of these charges 
for all component fields are the same as in \cite{BR} and are given in the Table 1. 
\begin{table}\setlength{\extrarowheight}{5pt}
\begin{center}\begin{tabular}{||l||lll|l|llll||} \hline\hline
  & $\Phi_1$ & $\Phi_2$ & $\Phi_3$ & $A_{\mu}$ & $\lambda_1$ & $\lambda_2$ & $\lambda_3$ & $\lambda_4$ \\
\hline\hline
$Q_1$ & $1$ & $0$ & $0$ & $0$ & $+\hf$ & $-\hf$ & $-\hf$ & $+\hf$ \\
$Q_2$ & $0$ & $1$ & $0$ & $0$ & $-\hf$ & $+\hf$ & $-\hf$ & $+\hf$ \\
$Q_3$ & $0$ & $0$ & $1$ & $0$ & $-\hf$ & $-\hf$ & $+\hf$ & $+\hf$ \\
\hline\hline
\end{tabular}\end{center}
\caption{\small Charges $Q_i$ of the component fields in the theory under the Cartan subgroup of
the $SU(4)_R$. }
\end{table}
These values are easy to derive from the fact that the integral of the superpotential of the $\cN=4$ SYM
\be
\int d^2 \theta \, {\cal W}_{\cN=4}\, =\,\int d^2 \theta \, i\, \Tr (\Phi_1\Phi_2\Phi_3 -\Phi_1\Phi_3\Phi_2)
\label{superpot}
\ee
is invariant under the action of each of these Cartan generators on the superfields $\Phi_i$
\be
\Phi_1 \, \to \, e^{i\phi_1}\, \Phi_1 \ , \qquad
\Phi_2 \, \to \, e^{i\phi_2}\, \Phi_2 \ , \qquad
\Phi_3 \, \to \, e^{i\phi_3}\, \Phi_3 
\label{sucharges}
\ee
This implies that the Grassmann $\cN=1$ superspace coordinate $\theta_\alpha$ is charged under
these transformations with $Q^{\theta}=(\hf,\hf,\hf).$
The charges of the scalar fields $\Phi_i$ are precisely the same as of their parent superfields
$\Phi_i$ in \eqref{sucharges} and the charges of the fermions $\lambda_A$ in the Table 1 are read from
Eq.~\eqref{sucharges} keeping in mind $\Phi_i(x,\theta) =\Phi_i(x) + \theta \cdot \lambda_i(x) + \ldots$.
The gauge field $A_\mu$ is neutral.\footnote{The fourth fermion $\lambda_4$ is the $\cN=1$ superpartner 
of $A_\mu$. It's charge is read off the invariance of the gauge kinetic term, $\int d^2 \theta WW$,
where $W_\alpha=\lambda_{4\alpha}+\ldots$ is the usual field-strength chiral superfield.}

The Lagrangian of the deformed theory follows from the component Lagrangian of the $\cN=4$ SYM
and the star-product definition \eqref{stardef}. We have:
\bea
\nonumber
&&{\cal{L}} = \,{1 \over g^2}\, \Tr \Bigg( {1 \over 4}
F^{\mu \nu}F_{\mu \nu} +
(D^\mu \bar \Phi^i ) (D_\mu \Phi_i  )
- {1\over 2} [\Phi_i,\Phi_j]_{C_{ij}}[\bar \Phi^i,\bar \Phi^j]_{C_{ij}}
+{1\over 4}[\Phi_i,\bar \Phi^i][\Phi_j,\bar \Phi^j] \\
&&+ \l_{A} \s^{\mu} D_{\mu} \bar\l^A
- i [\l_4,\l_i]_{B_{4i}}\bar \Phi^i+i[\bar\l^4,\bar \l^i]_{B_{4i}}\Phi_i
+{i \over 2}\epsilon^{ijk}[\l_i,\l_j]_{B_{ij}}\Phi_k
+{i \over 2}\epsilon_{ijk}[\bar\l^i,\bar\l^j]_{B_{ij}}\bar \Phi^k \Bigg)
\nonumber \\
\label{Ldef}
\eea
This Lagrangian contains only ordinary products between the fields; all modifications due to the
star-product \eqref{stardef} are assembled in \eqref{Ldef} into the deformed commutators of 
scalars $\Phi_i,$ $\bar \Phi^i$ and fermions $\lambda_A,$ $\bar \lambda^A.$ These deformed commutators
are 
\be
[\Phi_i,\Phi_j]_{C_{ij}}  \,:=\,
e^{ i C_{ij} }\,\Phi_i\Phi_j  -\,
e^{ -i C_{ij} }\, \Phi_j\Phi_i \ , \qquad i,j=1,2,3  \label{comC}
\ee
\be
[\lambda_A, \lambda_B]_{B_{AB}} \,:=\,
e^{ i  B_{AB} }\,\l_A \l_B  -\,
e^{ -i  B_{AB} }\, \l_B \l_A  \ , \qquad A,B=1,..,4 \label{comB}
\ee
Deformed commutators for $\bar \Phi$ and $\bar \lambda$ fields defined in the same way
as in \eqref{comC}-\eqref{comB}, and we note that the commutator $[\Phi_i,\bar \Phi^i]$
in \eqref{Ldef} is undeformed.
The matrices $C$ and $B$ are the same as in Ref.~\cite{BR}, they read
\be
C\, =\, \pi \,
\begin{pmatrix}
0 & -\gamma_3 & \gamma_2 \\
\gamma_3 & 0 & -\gamma_1 \\
-\gamma_2 & \gamma_1 & 0
\end{pmatrix}
\ , \quad
B\, =\, \pi \,
\begin{pmatrix}
0 & -\hf(\gamma_1+\gamma_2) & \hf(\gamma_3+\gamma_1) & \hf(\gamma_2-\gamma_3)\\
\hf(\gamma_1+\gamma_2) & 0 & -\hf(\gamma_2+\gamma_3) & \hf(\gamma_3-\gamma_1) \\
-\hf(\gamma_3+\gamma_1) & \hf(\gamma_2+\gamma_3) & 0 & \hf(\gamma_1-\gamma_2)\\
-\hf(\gamma_2-\gamma_3) & -\hf(\gamma_3-\gamma_1) & -\hf(\gamma_1-\gamma_2) & 0 
\end{pmatrix}
\label{CBmatr}
\ee
We see that the whole effect of the 3-parameter deformation is contained in
these matrices which introduce the appropriate phases into the 4-scalar and the Yukawa interactions
of the deformed theory \eqref{Ldef}. It is important to note that the induced phases of the fermions
(determined by the matrix $B$) are different from those of the scalars (in $C$). 
Also the ranks of $B$ and $C$ are different, the matrix $B$ introduces phases 
to the Yukawa interactions involving all the fermions, including the gaugino $\lambda_4$. 
The Lagrangian \eqref{Ldef} incorporates correctly the four-scalar interactions written down
in \cite{Frolov,FRT}. In addition to these, Eqs.~\eqref{Ldef}, \eqref{CBmatr} 
give the precise form of the interactions with fermions which are required for the instanton
calculations in the present paper. 

For a special case of all $\gamma_i$ being equal, the matrices 
$B$ and $C$ coincide with each other giving the same phase factors for scalars and fermions.
In this case, the gauge theory is $\cN=1$ supersymmetric 
and is dual to the supergravity solution of Lunin and Maldacena \cite{LM}. 
In the general case of unequal deformations $\gamma_i$, the fermion and scalar phases differ and the
gauge theory is non-supersymmetric.

Finally,
we need to make sure that the deformed gauge theory defined by 
Eqs.~\eqref{Ldef}, \eqref{CBmatr} is exactly marginal in the large $N$ limit. In general,
this would be a non-trivial task since the theory is not supersymmetric and one cannot use
the approach of Leigh and Strassler \cite{LS} to establish the required conformal invariance.
Instead the marginality of the theory follows from the use of the star-product.
It is known \cite{Filk} that the Moyal star-products used in the formulation of the noncommutative 
field theory do not affect the large $N$ perturbation theory. More precisely, the planar diagrams 
of the theories with and without the star-products can differ only by an overall phase-factor
which depends only on the external lines. This argument essentially uses only the associativity 
property of the star-product and it also applies to our choice \eqref{stardef}, see section \textbf{3.2} of
Ref.~\cite{VVK} for more detail. 
This implies that all planar perturbative contributions to the beta-functions and anomalous dimensions in
our deformed theory are proportional to those in the conformal $\cN=4$
theory, and hence vanish. Thus, the deformed theory \eqref{Ldef}, \eqref{CBmatr} is conformal
in the large $N$ perturbation theory.

The deformed theory \eqref{Ldef}, \eqref{CBmatr} is an interesting field theory on its own right.
It is a non-supersymmetric theory which fully inherits the remarkable structure of the 
large $N$ perturbation theory of the superconformal $\cN=4$ SYM. In Ref.~\cite{BDS} it was argued
that in the $\cN=4$ SYM the Maximally-Helicity-Violating (MHV) $n$-point amplitudes
have an iterative structure, such that the
kinematic dependence of all higher-loop
MHV amplitudes can be determined from the known one-loop results.
It then follows \cite{VVK} that 
the same must be true for the planar MHV amplitudes of the deformed theory. This is  a
consequence of the fact that the deformations were introduced via the star-product of the
type \eqref{stardef}. It is remarkable that such an iterative structure of the multi-loop
amplitudes can hold in a non-supersymmetric theory.

\section{Instanton effects}  \label{sec:insts}

Instantons in the deformed $\cN=4$ gauge theory have been discussed in detail in
Ref.~\cite{GK}. We refer the reader to this reference and summarise here only the key
points. The instanton configuration is defined to the leading order in the Yang-Mills coupling $g$,
and satisfies the following equations for the gauge field, 
\be
F_{mn} = {}^* F_{mn}
\label{sdeq}
\ee
fermions, 
\be
\Dbarslash^{\dalpha \alpha} \lambda_{\alpha}^A =\ 0
\label{fzerms}
\ee
and scalars,
\be
\D^2 \Phi^{AB}\ =\ \sqrtwo
i\,(\,e^{i B_{AB}}\lambda^A  \lambda^B\,-\,e^{-i B_{AB}}\lambda^B \lambda^A\,)
\label{Higgseq}
\ee
Here $\Dbarslash^{\dalpha \alpha}= D^\mu \sigmabar_\mu^{\dalpha \alpha}$ and
$\D^2= D^\mu D_\mu$ where $D_\mu$ is the covariant derivative in the
instanton  background $A_\mu$. The matrix $B$ is given in \eqref{CBmatr}.

There are $8kN$ fermionic solutions of
\eqref{fzerms} in the $k$-instanton background. 
16 of these solutions correspond to $2\cN=8$ supersymmetric and 
$2 \cN=8$ superconformal fermion zero modes of the original $\cN=4$ gauge theory.
In the $\cN=4$ SYM these 16 fermion zero modes are exact.
In our deformed theory supersymmetry is lost and all of the fermion zero modes are lifted
in the instanton action as will be seen shortly.

The scalar field equation \eqref{Higgseq} follows from the Yukawa interactions\footnote{The 
four-scalar interactions in \eqref{Ldef} do not enter the leading order in $g$ instanton
construction.}
in \eqref{Ldef} and is written in the basis $\Phi^{AB}=-\Phi^{BA}$ for the scalar fields.
This representation is related as follows to the usual basis $\Phi_i$ used in \eqref{Ldef}
(see \cite{GK})
\bea
\Phi_1 = \, \frac{1}{\sqrt{2}} (\phi^1 +i\phi^2) &=& 2 \,\bar{\Phi}^{23} = \,2\,\Phi^{14}
\nonumber\\
\Phi_2 = \, \frac{1}{\sqrt{2}} (\phi^3 +i\phi^4) &=& 2 \,\bar{\Phi}^{31} = \,2\,\Phi^{24}
\label{dictphi}\\
\Phi_3 = \, \frac{1}{\sqrt{2}} (\phi^5 +i\phi^6) &=& 2 \,\bar{\Phi}^{12} = \,2\,\Phi^{34}
\nonumber
\eea
The instanton configuration is the solution of the defining equations \eqref{sdeq}-\eqref{Higgseq},
and it is used to construct the semi-classical instanton integration measure. This measure 
is an integral over the instanton collective coordinates and it determines
instanton contributions to the path integral. The general multi-instanton measure was
constructed in \cite{MO3} for the $\cN=4$ SYM and generalised in \cite{GK} to account for the
supersymmetry preserving $\beta$-deformations. The result of \cite{GK} can be now straightforwardly
adopted to the case of non-supersymmetric $\gamma_i$ deformations. We will concentrate here on the
simplest case of the single-instanton measure. The multi-instanton measure is a straightforward
generalisation of thereof along the lines of \cite{GK,MO3}.

The 1-instanton measure of the $\gamma_i$-deformed theory reads (cf. Eq.~(5.2) of Ref.~\cite{GK}):
\bea
\int d\mu\,e^{-S_{\rm 1-inst}} \, =\,
{2^{-31}\pi^{-4N-5} g^{4N}\over(N-1)!(N-2)!}
\,\int d^4 x_0\, d\rho \, d^6 \chi \, 
\prod_{A=1}^{4} d^2\xi^A\, d^2\bar\eta^A\,
d^{(N-2)}\nu^A\, d^{(N-2)}\bar\nu^A
\nonumber \\
\rho^{4N-7}\, \exp\big[-{8\pi^2\over g^2} +i\theta- 2\rho^2 \chi^a \chi^a
+{4\pi i \over g}\,\chi_{AB}\Lambda_{AB}\big]
\label{1instpf}
\eea
The integral above is over the bosonic and fermionic (Grassmann) collective coordinates of the instanton.
The fermionic ones comprise $4(N-2)$ parameters $\nu^A_i$ (where $i=1,\ldots N-2$), 
$8$ supersymmetric coordinates $\xi^A_\alpha$ and
$8$ superconformal modes $\bar\eta^A_{\dot\alpha}$ (where $\alpha=1,2$ and $\dot\alpha=1,2$).
Bosonic collective coordinates include the instanton position $x_0^\mu$, the scale-size $\rho$
and the 6 additional variables $\chi^a$ which are coupled to fermion modes in the instanton action
in the exponent in \eqref{1instpf}.
The variables $\chi^a$ or $\chi_{AB}$ transform in the vector
representation of the $\SO(6)\cong\SU(4)$ R-symmetry and is subject to
the reality condition 
$\chi_{AB}^\dagger = \, \hf
\epsilon^{}_{ABCD}\chi^{}_{CD}.$

Finally $\Lambda_{AB}$ in the instanton action in \eqref{1instpf} is a fermionic bilinear 
defined as
\be
\Lambda_{AB} = \, {1 \over 2\sqrt{2}}\, \sum_{i=1}^{N-2}\left(e^{i B_{AB}}\,\bar\nu^A_i \nu^B_i 
-e^{-i B_{AB}}\,\bar\nu^B_i \nu^A_i\right) 
+ i 8\sqrt{2}\, \sin(B_{AB}) \,\left(\rho^2 \bar\eta^A \cdot \bar\eta^B +\xi^A \cdot\xi^B\right)
\label{Lam1inst}
\ee
The $4 \times 4$ antisymmetric matrix
$B_{AB}$ was defined in \eqref{CBmatr}.
The fact that the instanton action in \eqref{1instpf} depends on all of the fermionic collective coordinates
(through $\Lambda_{AB}$) implies that they are lifted. This is to be expected in the non-supersymmetric
theory.

 Following the approach of \cite{GK}
 we proceed by integrating out fermionic collective coordinates
 $\nu_i^A$ and $\bar\nu_i^A$ from the instanton partition function \eqref{1instpf}.
 For each value of $i=1,\ldots,N-2$ this integration gives a factor of 
 \be
 \left({4 \pi \over g}{1 \over \sqrt{2}}\right)^4 \, 
 {\rm det}_4 \left(e^{i B_{AB}}\,\chi_{AB}\right)
 \label{detbchi}
\ee
The determinant above can be calculated directly. It will be useful
to express the result in terms of the three complex variables
$X_i$ which are defined in terms of $\chi_{AB}$ in the way analogous to
Eqs.~\eqref{dictphi}:
\bea
X_1 = \, \chi^1 +i\chi^2 &=& 2\sqrt{2} \,{\chi}^\dagger_{23} = \,2\sqrt{2}\,\chi_{14}
\nonumber\\
X_2 = \, \chi^3 +i\chi^4 &=& 2\sqrt{2} \,{\chi}^\dagger_{31} = \,2\sqrt{2}\,\chi_{24}
\label{dictchi}\\
X_3 = \,  \chi^5 +i\chi^6 &=& 2\sqrt{2} \,{\chi}^\dagger_{12} = \,2\sqrt{2}\,\chi_{34}
\nonumber
\eea
In terms of these degrees of freedom, the determinant takes the form
\be
{\rm det}_4 \left(e^{i B_{AB}}\,\chi_{AB}\right) =
\begin{vmatrix}
0 & X_3^\dagger\, e^{-\ft{i\pi}{2}(\gamma_1+\gamma_2)} & -X_2^\dagger\, e^{\ft{i\pi}{2}(\gamma_3+\gamma_1)} & X_1\, e^{\ft{i\pi}{2}(\gamma_2-\gamma_3)}\\
-X_3^\dagger\, e^{\ft{i\pi}{2}(\gamma_1+\gamma_2)} & 0 &  X_1^\dagger\, e^{-\ft{i\pi}{2}(\gamma_2+\gamma_3)} & X_2\, e^{\ft{i\pi}{2}(\gamma_3-\gamma_1)}\\
X_2^\dagger\, e^{-\ft{i\pi}{2}(\gamma_3+\gamma_1)} & -X_1^\dagger\, e^{\ft{i\pi}{2}(\gamma_2+\gamma_3)} & 0 & X_3\, e^{\ft{i\pi}{2}(\gamma_1-\gamma_2)}\\
-X_1\, e^{-\ft{i\pi}{2}(\gamma_2-\gamma_3)} & -X_2\, e^{-\ft{i\pi}{2}(\gamma_3-\gamma_1)} & -X_3\, e^{-\ft{i\pi}{2}(\gamma_1-\gamma_2)} & 0 
\end{vmatrix} 
\ee
It is evaluated to give
\bea
{\rm det}_4 \left(e^{i B_{AB}}\,\chi_{AB}\right)
&=&
{1 \over 64} \, (|X_1|^2 + |X_2|^2 + |X_3|^2)^2
\label{defdet} \\
&-& \, {1\over 16}\, \sin^2 (\pi \gamma_3)\,|X_1|^2|X_2|^2 
\,-\,{1\over 16}\, \sin^2 (\pi \gamma_1)\,|X_2|^2|X_3|^2 
\,-\,{1\over 16}\, \sin^2 (\pi \gamma_2)\,|X_3|^2|X_1|^2 
\nonumber
\eea
We note that the expression above depends only on the three absolute
values of $|X|$ and is independent of the three angles. 
We can further change variables as follows:
\be
\label{mudefs}
|X_i| =\, r\, \mu_i \ , \qquad \sum_{i=1}^3 \mu_i^2 =\,  1
\ee
and write
\bea
 \left({4 \pi \over g}{1 \over \sqrt{2}}\right)^4 \, 
 {\rm det}_4 \left(e^{i B_{AB}}\,\chi_{AB}\right)=\, 
  \left( {\pi \over g}\right)^4 \, r^4 \, \Big( 1&&
  \, -\, 4\, \sin^2 (\pi \gamma_3)\, \mu_1^2\mu_2^2\\\nonumber
 && \, -\, 4\, \sin^2 (\pi \gamma_1)\, \mu_2^2\mu_3^2
  \, -\, 4\, \sin^2 (\pi \gamma_2)\, \mu_3^2\mu_1^2
  \Big) 
\label{e58}
\eea

In summary after integrating out all of the $\nu$ and $\bar\nu$ fermionic
collective coordinates we find the following generic instanton factor in the measure:
\bea
\label{OneIM}
{\cal F}_{\rm inst} &:=& e^{-{8\pi^2\over g^2}+i\theta}\,
\left(1\, 
 \, -\, 4\, \sin^2 (\pi \gamma_3)\, \mu_1^2\mu_2^2
  \, -\, 4\, \sin^2 (\pi \gamma_1)\, \mu_2^2\mu_3^2
  \, -\, 4\, \sin^2 (\pi \gamma_2)\, \mu_3^2\mu_1^2
\right)^{N-2}\\
&\equiv& e^{2\pi i\tau_0}\left( 1\, -\, \cQ(\mu_i,\gamma_i)\right)^{N-2} \nonumber
\eea
This factor is integrated over the $\adss$ space spanned by $x_0^\mu$, $\rho$ and the five angles
of $\chi^a$
\be
\int\,  d^4 x_0 \,\, {d\rho \over \rho^5} \,\, d^5 {\hat\chi}\, =\,
(2\pi)^3 \,  \int\,  d^4 x_0 \,\, {d\rho \over \rho^5} \,\, d\mu_1\,d\mu_2\,d\mu_3\, 
\delta(\mu_1^2+\mu_2^2+\mu_3^2-1)
\ee
exactly as in \cite{GK,MO3}. As we are interested in the limit
$N\rightarrow\infty$ we can rewrite Eq.~\eqref{OneIM} as a total exponent and
evaluate the integrals over $\mu_i$ via a saddle-point
approximation,
\bea
\nonumber
\int_{\mu_i}\, e^{2\pi i\tau_0}\left( 1\, -\, \cQ(\mu_i)\right)^{N-2}
&=& \int_{\mu_i}\,\exp{\Bigl( 2\pi i\tau_0 +(N-2)\,\log {\bigl(1\, -\, \cQ(\mu_i)\bigr)}\Bigr)} \\
&\approx& \exp{\bigl( 2\pi i\tau_0 \,-\,N\,\cQ(\mu_i\vert_{\rm{saddle}})\bigr)} \label{saddleOneI}
\eea
This method selects the dominant value of the function $\cQ(\mu_i)$ to be $\cQ(\mu_i\vert_{\rm{saddle}})\sim
{1\over N}$ and has therefore allowed us to expand the log to leading power in
$\cQ$ in the last line.

What we have calculated so far is a large-N expression for the characteristic instanton factor 
\be
\label{Ffactor}
{\cal F}_{\rm inst}\, =\, \exp{\bigl( 2\pi i\tau_0 \,-\,N\,\cQ(\mu_i\vert_{\rm{saddle}},\gamma_i)\bigr)}
\ee
This factor arises in an instanton calculation of a generic correlation function in gauge theory.
When applied to the calculation of Yang-Mills correlators involving operators which are dual to the supergravity
fields, the instanton result in gauge theory must match with the corresponding D-instanton contribution in string theory.
This means that the characteristic factor \eqref{Ffactor} due to the Yang-Mills instanton must correspond to
$\exp\bigl( 2\pi i\tau\bigr),$ where $\tau$ is the dilaton-axion field in 
dual string theory.\footnote{More detail about instanton and D-instanton contributions to the string effective action can be found
in \cite{GK,MO3,BG,BGKR}.}
By matching exponents we read off the instanton prediction for the
dilaton-axion field:
\be
\label{OneITau}
\tau = \tau_0 - {N \over 2\pi i}\left( 4\, \sin^2 (\pi \gamma_3)\, \mu_1^2\mu_2^2
  \, +\, 4\, \sin^2 (\pi \gamma_1)\, \mu_2^2\mu_3^2
  \, +\, 4\, \sin^2 (\pi \gamma_2)\, \mu_3^2\mu_1^2 \right)
\ee
We note that
this semi-classical field theory result is valid for any value of the parameters $\gamma_i$
and, as such, can be interpreted \cite{CK} as a (weak-coupling) prediction for the $\tau$ field in the exact string theory
background.

The supergravity regime is reached in the limit of $\gamma_i \ll 1$
which gives:
\be
\label{OneIsugra}
\tau \, \rightarrow\, \tau_0 + {2N\pi i}\left( \gamma_3^2 \, \mu_1^2\mu_2^2
  \, +\, \gamma_1^2 \, \mu_2^2\mu_3^2
  \, +\, \gamma_2^2 \, \mu_3^2\mu_1^2 \right)
\ee
This precisely matches with the Frolov's three parameter supergravity solution
\eqref{threepar} for the dilaton-axion field:
\bea
\label{3tau}
\tau &=& ie^{-\phi}+C\\
     &=& ie^{-\phi_0}\Bigl(1\,+\, \hg_3^2 \, \mu_1^2\mu_2^2
                           \, +\, \hg_1^2 \, \mu_2^2\mu_3^2
                           \, +\, \hg_2^2 \, \mu_3^2\mu_1^2 \Bigr)^{1/2}+C^0\\\nonumber
     &=& \tau_0\, +\, {ie^{-\phi_0}\over 2}\left(\hg_3^2 \, \mu_1^2\mu_2^2
                           \, +\, \hg_1^2 \, \mu_2^2\mu_3^2
                           \, +\, \hg_2^2 \, \mu_3^2\mu_1^2 \right)
\eea
where the deformation parameters are
     $\hg_i^2= N\,g^2\,\gamma_i^2$ and one identifies the coordinates on the
     deformed supergravity $\tilde S^5$ sphere with the $\chi$-collective
     coordinates of the instanton.

It is clear in the above that an analogous calculation for the case
of one \emph{anti}-instanton would yield the same type of gauge/supergravity
matching for the conjugate parameter $\bar \tau$. One can also extend this
calculation to include the multi-instanton sectors, as in 
\cite{GK,MO3}. In the large
$N$ limit the partition function in the $k$-instanton sector is:
\be
\int\, d\mu^k_{\rm inst}\,e^{-\Skinst}\ {=}\ {\sqrt{Ng^2} \over
2^{33}\pi^{27/2}}\,{k\over g^2}^{-7/2}\sum_{d\vert k}{1\over d^2}
\int\,
{d^4x\,d\rho\over\rho^5}\, d^5\hat\Omega \prod_{A=1,2,3,4}d^2\xi^A 
d^2\bar\eta^A\\ e^{2\pi ik \tau}
\label{kIpartition}
\ee
where $\tau$ is given by the same Eq.~\eqref{OneITau}. 

\section{Complex $\beta$ deformations} \label{sec:gencompl}

In this section we consider the more general case 
of marginal deformations with complex values of the deformation parameters $\beta_i \in \bigC$
\be
\beta_1 \, =\, \gamma_1\, +\, i\, \sigma_1 \ , \quad \beta_2 \, =\, \gamma_2\, +\, i\, \sigma_2 \ ,
\quad \beta_3 \, =\, \gamma_3\, +\, i\, \sigma_3
\ee 
The supergravity solution corresponding to this case was obtained in \cite{Frolov} by performing three consecutive 
$STsTS^{-1}$ transformations (where S is the S-duality) acting on the three natural tori of $S^5$.
 This family of solutions is expected to be dual to a
deformed Yang-Mills theory with three complex deformation parameters.

We will first explain how to extend the instanton calculation on the gauge theory
side from real to complex $\beta_i$-deformations. We will carry out this calculation
for arbitrary (not necessarily small) values of the deformation parameter $\beta_i \in \bigC$.
 The main result of this section is the instanton prediction for 
the dilaton-axion field $\tau$. We will show that in the limit of small $\beta_i$
it will match precisely with the $\tau$ field of Frolov's supergravity 
dual \cite{Frolov}. As before, the small-$\beta_i$ limit is required to ensure the 
validity of the supergravity approximation to full string theory. 

We now need to specify the deformed gauge theory. 
The absence of supersymmetry and the complex-valuedness of the deformations $\beta_i$
make it difficult. 
It is not entirely clear how to
uniquely define this theory and, more importantly, how to guarantee its marginality
in the large $N$-limit.\footnote{The absence of supersymmetry prevents one from using the 
Leigh-Strassler approach \cite{LS} in terms of conformal constraints, while the
complex-valuedness of the deformation parameters makes it difficult to use the star-product
formulation.}
Fortunately,
the instanton calculation which we are about to present does not require the full knowledge of the
gauge-theory Lagrangian, beyond its gauge and Yukawa interactions specified below.

The instanton configuration at the leading order in weak coupling is defined as in
equations \eqref{fzerms}-\eqref{Higgseq}
with the scalar field equation \eqref{Higgseq}
taking the form:
\bea
&\D^2 \Phi^{AB}\ =\ \frac{h}{g} \, \sqrtwo
i\, (\,e^{i\pi{\rm B}^{AB}}\, \lambda^A  \lambda^B\,-\,e^{-i\pi{\rm B}^{AB}}\,\lambda^B \lambda^A\,)
\qquad  \ ,
\label{Higgseq2}
\eea
Here
$B_{AB}$ is a complex-valued matrix 
obtained from the one in \eqref{CBmatr} by
the substitution $\gamma_i \rightarrow \beta_i$. 
The factor of $h/g$ on the right hand side of \eqref{Higgseq2} accounts for the change of
the coupling constant from $g$ to $h$ in the Yukawa couplings, where $h$ is an new complex parameter.
We note that the resulting instanton configuration depends on $h$ holomorphically.\footnote{At leading order 
in $g$ the dependence on $h^*$ 
can come only through the equation conjugate to \eqref{Higgseq2}, which involves 
anti-fermion zero modes $\bar\lambda$ on the right hand side. These are vanishing 
in the instanton background. 
It is clear then that the anti-instanton configuration, will depend on $h^*$ and not on $h$.}

Following the approach of section \textbf{3} we integrate out fermionic collective coordinates
$\nu_i^A$ and $\bar\nu_i^A$.
For each value of $i=1,\ldots,N-2$ this integration gives a factor of the
determinant \eqref{detbchi} times an appropriate rescaling by $h/g$. 
We find
\be
 \left({1 \over g}\right)^4 \, 
 {\rm det}_4 \left(e^{i\pi {\rm B}^{AB}}\,\chi_{AB}\right) \, \longrightarrow\,
  \left({1 \over g}\right)^4 \, \left({h \over g}\right)^2\,
 {\rm det}_4 \left(e^{i\pi {\rm B}^{AB}}\,\chi_{AB}\right) \ .
  \label{e58h}
\ee
After evaluating this determinant,
the result for the  characteristic instanton factor in the large-N
limit is:
\be
{\cal F}_{\rm inst} \, =\, \exp\left[ 2\pi i \tau_0\, +\, 2N\log \left({h \over g}\right)\,
+\, N \log \left(1\, -\, \cQ(\mu_i,\beta_i) \right) \right ]
\label{Fcomp2},
\ee
where $\cQ(\mu_i,\beta_i)$ is the same function as before, but with 
the complex $\beta_i$ parameters in place of real $\gamma_i$,
\be
\cQ(\mu_i,\beta_i) \, =\,
4\,\left( \sin^2 (\pi \beta_3)\, \mu_1^2\mu_2^2
  \, +\,  \sin^2 (\pi \beta_1)\, \mu_2^2\mu_3^2
  \, +\,  \sin^2 (\pi \beta_2)\, \mu_3^2\mu_1^2\right)
\ee

By taking the small deformation limit, $|\beta_i|^2\ll 1$,
appropriate for comparison with the supergravity solution, we find                                                         
\bea
{\cal F}_{\rm inst}\, =\, \exp \Big[2\pi i \tau_r \,-\,4\pi^2\,N\, 
((\gamma_1^2-\sigma_1^2+ 2i\gamma_1\sigma_1 )\mu_2^2\mu_3^2+
(\gamma_2^2-\sigma_2^2+ 2i\gamma_2\sigma_2 )\mu_1^2\mu_3^2  \nonumber \\
+(\gamma_3^2-\sigma_3^2+ 2i\gamma_3\sigma_3 )\mu_2^2\mu_1^2
)\Big],
\label{Fcom}
\eea
Here $\tau_r$ is the constant which has the meaning of the `renormalised'
 Yang-Mills coupling as in section~\textbf{8}
of \cite{GK} and in \cite{DHK}. It is defined via
\be
\tau_r \, :=\, \tau_0 \, -\, {i N \over \pi} \, \log {h \over g}
\ee

The dilaton and axion field components of the Frolov's supergravity dual with three complex deformations
are given by \cite{Frolov}:
\bea
&e^\phi \, = \, e^{\phi_0}\,G^{1/2}H \, ,\qquad 
&C\, =\, C ^0 \,+ \,e^{-\phi_0} \,H^{-1}Q \, ,
\label{hatdefs}
\eea
where the expressions for the functions $G$, $H$ and $Q$ can be found in the Appendix B of \cite{Frolov}.
By employing these expressions and \eqref{hatdefs} one can easily calculate the axion-dilaton field
for the case of complex deformations. The result thus obtained reads
\bea
e^{2 \pi i \tau}\,& =&\, e^{2 \pi i (i e^{-\phi}+C)}\, =\, 
\exp  \Big[-2 \pi e^{-\phi_0}[1+\hf({\hat\gamma_1}^2-{\hat\sigma_1}^2)\mu_2^2\mu_3^2
+\hf({\hat\gamma_2}^2-{\hat\sigma_2}^2)\mu_1^2\mu_3^2+
\hf({\hat\gamma_3}^2-{\hat\sigma_3}^2)\mu_2^2\mu_1^2]\, \cr
&+&\, 2\pi i(C^0 + e^{-\phi_0}({\hat\gamma_1}{\hat\sigma_1}\mu_2^2\mu_3^2+
{\hat\gamma_2}{\hat\sigma_2}\mu_1^2\mu_3^2+
{\hat\gamma_3}{\hat\sigma_3}\mu_2^2\mu_1^2 ))\Big] \label{fin-al}
\eea
By making the identification
\be
\hat\gamma_i=g_r\sqrt{N}\gamma_i \ , \qquad \hat\sigma_i=-g_r\sqrt{N}\sigma_i\ ,
\qquad \tau_r = ie^{-\phi_0} + C^0
\ee
one can immediately see this supergravity result is in
 perfect agreement 
 with our field theory prediction \eqref{Fcom}.

\bigskip
\bigskip

\centerline{\bf Acknowledgements}

We thank Sergey Frolov, Chong-Sun Chu and Stefano Kovacs for useful discussions and comments.
Our research 
is supported by PPARC through a Studentship, a Postdoctoral Fellowship and a
Senior Fellowship.


\end{document}